\documentclass[11pt]{article}
\usepackage{fa2025}
\usepackage{amsmath}
\usepackage{cite}
\usepackage{url}
\usepackage{graphicx,subcaption}
\usepackage{color}
\usepackage{siunitx}
\usepackage[utf8]{inputenc}
\title{Enhancing photogrammetry reconstruction for HRTF synthesis via a Graph Neural Network}

\threeauthors
  {Ludovic Pirard} {Dyson School of Design Engineering, Imperial College London, London, UK \\ {\tt l.pirard@imperial.ac.uk}}
    {Katarina Poole} {Dyson School of Design Engineering, Imperial College London, London, UK \\ {\tt katarina.poole@imperial.ac.uk}}
  {Lorenzo Picinali} {Dyson School of Design Engineering, Imperial College London, London, UK \\ {\tt l.picinali@imperial.ac.uk}}
\multauthor
{Ludovic Pirard$^{1*}$ \hspace{1cm} Katarina Poole$^1$ \hspace{1cm} Lorenzo Picinali$^1$} { %\bfseries{Fourth author$^3$ \hspace{1cm} Fifth author$^2$ \hspace{1cm} Sixth author$^1$}\\
  $^1$ Dyson School of Design Engineering, Imperial College London, London, United Kingdom\\
\correspondingauthor{l.pirard@imperial.ac.uk}{Pirard et al.}
}

\begin{document}

\maketitle
\begin{abstract}
Traditional Head-Related Transfer Functions (HRTFs) acquisition methods rely on specialised equipment and acoustic expertise, posing accessibility challenges. Alternatively, high-resolution 3D modelling offers a pathway to numerically synthesise HRTFs using Boundary Elements Methods and others. However, the high cost and limited availability of advanced 3D scanners restrict their applicability. Photogrammetry has been proposed as a solution for generating 3D head meshes, though its resolution limitations restrict its application for HRTF synthesis. To address these limitations, this study investigates the feasibility of using Graph Neural Networks (GNN) using neural subdivision techniques for upsampling low-resolution Photogrammetry-Reconstructed (PR) meshes into high-resolution meshes, which can then be employed to synthesise individual HRTFs. Photogrammetry data from the SONICOM dataset are processed using Apple Photogrammetry API to reconstruct low-resolution head meshes. The dataset of paired low- and high-resolution meshes is then used to train a GNN to upscale low-resolution inputs to high-resolution outputs, using a Hausdorff Distance-based loss function. The GNN’s performance on unseen photogrammetry data is validated geometrically and through synthesised HRTFs generated via Mesh2HRTF. Synthesised HRTFs are evaluated against those computed from high-resolution 3D scans, to acoustically measured HRTFs, and to the KEMAR HRTF using perceptually-relevant numerical analyses as well as behavioural experiments, including localisation and Spatial Release from Masking (SRM) tasks. 
\end{abstract}

\section{Introduction}\label{sec:introduction}

\subsection{Related Works}

Head-Related Transfer Functions are unique acoustic filters that characterise how sound waves from locations around the listener interact with their anatomy, including the shape, size, and position of the ears, as well as the head and torso \cite{engelSONICOMHRTFDataset2023}.

Generic HRTFs can be derived from average human morphology using mannequins such as the KEMAR \cite{gardner_hrtf_1995}, performing acoustic measurements or numerical synthesis from 3D meshes \cite{ziegelwanger_mesh2hrtf_2015}. While these are useful for universal consumer applications, they often fail to provide a sufficiently accurate spatial audio experience, resulting in front-back confusion and impaired elevation perception \cite{brinkmann_cross-evaluated_2019}.

Individual HRTFs are fundamental in immersive audio as they can result in enhanced rendering quality, higher sound localisation accuracy, and potentially also better spatial release from masking performances \cite{gonzalez-toledo_spatial_2024, picinali_system--user_2023, meyer_generalization_2025}.

Traditional individual HRTF acquisition methods require specialised equipment and expertise, which pose significant accessibility challenges \cite{pauwels_relevance_2023}. Alternatively, high-resolution 3D modelling provides a pathway to numerically synthesised HRTFs widely available using tools such as Mesh2HRTF \cite{ziegelwanger_mesh2hrtf_2015}. However, the limited availability and high cost of advanced 3D scanners restrict their applicability \cite{pollack_perspective_2022, ziegelwanger_calculation_2013}. 

Photogrammetry can be seen as an alternative; it is more accessible as well as affordable, and can be performed employing consumer equipment, e.g., a smartphone or a digital camera. Photogrammetry has been explored as a solution to obtain HRTFs via mesh reconstruction and HRTFs synthesis in previous works \cite{dellepiane_reconstructing_2008, pollack_perspective_2022, meshram_p-hrtf_2014}. The lack of resolution and ear details in the mesh reconstruction has limited the use of photogrammetry for personal HRTFs synthesis \cite{pollack_application_2023, pollack_combination_2024}.

\subsection{Research aims}
To address these limitations, this study investigates the feasibility of using Graph Neural Networks (GNN) \cite{hanocka_meshcnn_2019} implementing neural subdivision techniques for upsampling low-resolution photogrammetry-reconstructed (PR) meshes into high-resolution ones, which can then be employed to synthesise individual HRTFs. The objective is to improve the resolution and ear morphology details from photogrammetry-reconstructed meshes.

The overall goal is to create an accessible method for any user to obtain a personal HRTF that closely resembles what they could obtain from acoustically measured HRTF, offering more accurate spatial cues compared to a generic HRTF. 
%The ultimate objective is to design an automated pipeline that seamlessly transitions from photogrammetry to individual HRTFs. 

%
%\section{Paper Length AND File Size}
%The authors can submit for the EAA FA\conferenceyear\ Conference either a short paper from 2 to 4 pages or a long paper of maximum 8 pages, all written in English, and not previously published. Each length indicated is intended as overall and includes figures, tables, references and acknowledgments.\\\\
%Papers should be submitted as PDF files. The file size should not exceed 25 MB. Please compress images and figures as necessary before submitting.

\section{Methodology}\label{sec:methodology}

HRTF synthesis is first performed on raw photogrammetry-reconstructed meshes using Mesh2HRTF to establish a baseline and assess the limitations of photogrammetry for individual HRTF computation, without the use of neural networks. Both numerical and perceptual evaluations are conducted. Subsequently, a Graph Neural Network (GNN) is trained and evaluated to generate upsampled meshes. HRTFs are synthesised from these refined meshes and compared against the baseline results obtained from the raw photogrammetry data. Finally, numerical and perceptual evaluations are also carried out to assess the improvements achieved through refinement.

\subsection{Photogrammetry Reconstruction}

For each subject in the SONICOM dataset \cite{engelSONICOMHRTFDataset2023}, 72 images were captured at 5-degree intervals to achieve a full 360-degree representation. The photogrammetry data include high-resolution RGB photographs, depth maps, and gravity data, all acquired using an iPhone XS with a custom 3D-printed mirror bracket to use Apple's TrueDepth technology.

To evaluate the effectiveness of different photogrammetry reconstruction methods, multiple software solutions were tested, including Reality Capture, Agisoft Metashape, Autodesk Recap Photo, and Apple’s photogrammetry API. An informal assessment of mesh quality, resolution, and ear morphology details was conducted, revealing that Apple’s photogrammetry API produced the most accurate and visually consistent reconstructions. Consequently, this method was implemented using Swift within Xcode for batch processing.

The resulting 244 subject meshes are generated in the .stl format from the SONICOM dataset photogrammetry data. Each subject mesh obtained from photogrammetry has a corresponding high-resolution 3D scan acquired with an EXScan Pro. The main differences between the two meshes are the number of faces and vertices, which impacts the resolution of the meshes and the ear details. The photogrammetry mesh still represents the general shape of the head and ears but lacks ear features that represent individual ear morphology. 

\begin{figure}[ht]
    \centering
    \begin{subfigure}[t]{0.48\linewidth}        
        \includegraphics[width=\linewidth]{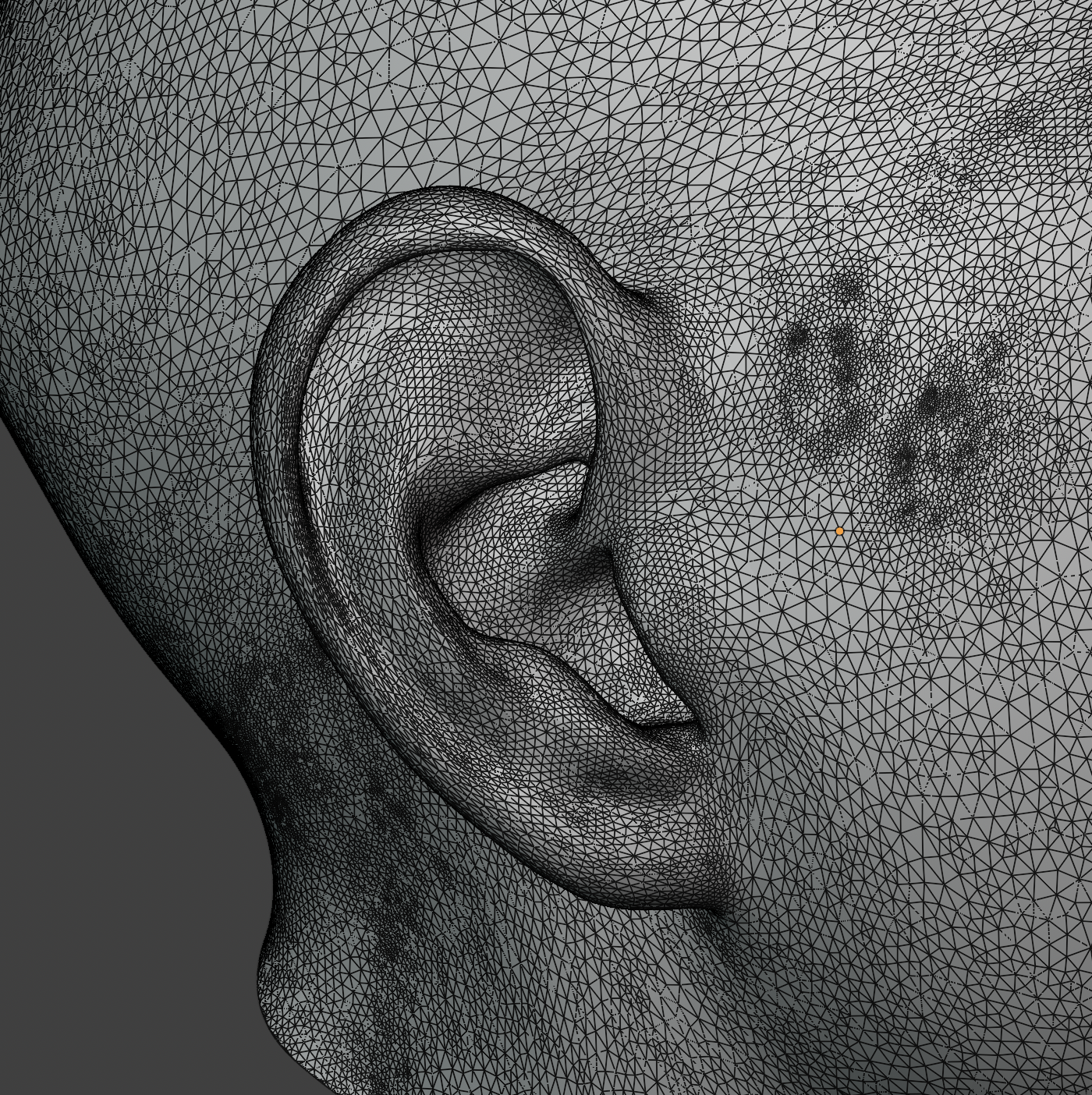}
        \caption{3D scan mesh : right ear}
        \label{fig:A}
    \end{subfigure}
    \hfill
    \begin{subfigure}[t]{0.48\linewidth}      
        \includegraphics[width=\linewidth]{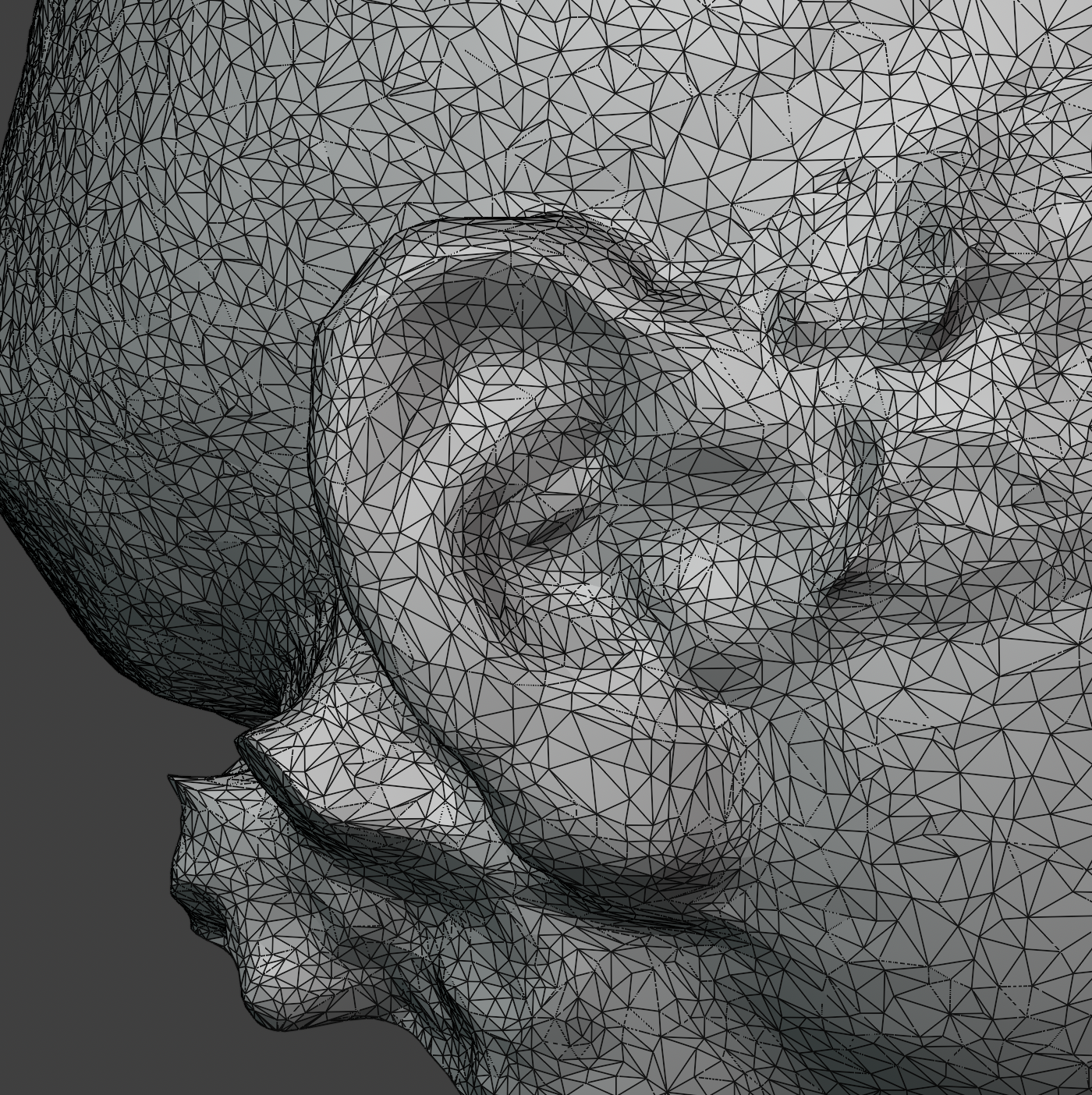}
        \caption{PR mesh using Apple Photogrammetry API : right ear}
        \label{fig:B}
    \end{subfigure}
    \caption{Same subject right ear meshes with different acquisition methods}
    \label{fig:PRRightEar}
\end{figure}

\subsection{HRTF synthesis}

Meshes of varying quality including raw photogrammetry reconstructions, GNN upsampled meshes, and high-resolution 3D scans are processed using Mesh2HRTF \cite{ziegelwanger_mesh2hrtf_2015} to generate simulated HRTFs with chosen locations that correspond to the SONICOM HRTF measurement setup. Prior to simulation, several post-processing steps are applied, including mesh alignment, beheading, clean-up, curvature-adaptive mesh grading, and the assignment of distinct mesh faces for the skin, right ear, and left ear.

Curvature-adaptive mesh grading refines the mesh resolution by increasing the density of elements in high-curvature regions, such as the pinnae, while reducing complexity in flatter areas \cite{palm_curvature-adaptive_2021}. This optimisation maintains essential geometric features for accurate acoustic diffraction modelling while minimising computational cost. Due to the high memory requirements of the boundary element method (BEM) used in Mesh2HRTF, simulations are performed on a High-Performance Computing (HPC) system. The HRTF synthesis process follows the guidelines provided in the Mesh2HRTF documentation and tutorials to ensure methodological consistency.

\subsection{Numerical and perceptual evaluation}

Synthesised HRTFs obtained from photogrammetry-reconstructed and GNN-upsampled meshes are evaluated numerically and perceptually against those computed from high-resolution scans, acoustically measured HRTFs, and the KEMAR HRTF.

The numerical evaluation is conducted using the Log-Spectral Distortion (LSD) metric \cite{hu_hrtf_2024}, which quantifies spectral differences between HRTFs in the frequency domain, as well as interaural cue variations, including Interaural Time Differences (ITDs) and Interaural Level Differences (ILDs). Numerical analysis is performed using the Spatial Audio Metrics 0.1.2, K. C. Poole, AXD, Imperial College London \url{https://github.com/Katarina-Poole/Spatial-Audio-Metrics}. 

\begin{figure}[ht]
    \centerline{{\includegraphics[width=8.5cm]{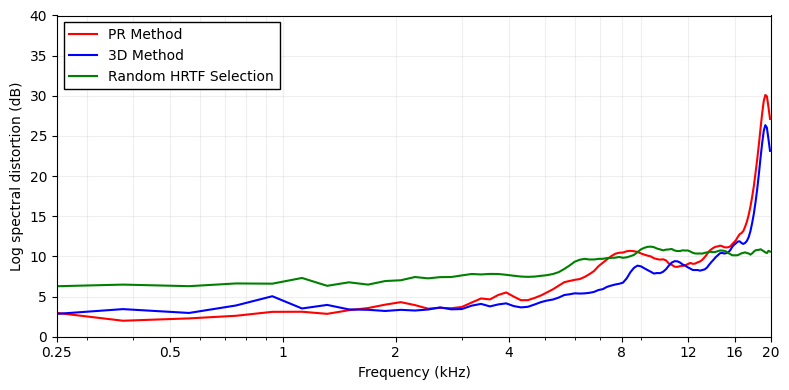}}}
    \caption{Average LSD comparison across 10 subjects : PR vs 3D vs Random HRTF selection against measured HRTF}
    \label{fig:LSDDif}
 \end{figure}

The average Log-Spectral Distortion (LSD) is computed across 10 subjects to compare HRTFs synthesised from photogrammetry-Reconstruction (PR) and high-resolution 3D scan meshes \cite{poole_extended_2025} against measured HRTFs. For each subject, LSD is calculated as the mean distortion between the synthesised and measured HRTFs for both ears, and then averaged across subjects. Results in Fig. \ref{fig:LSDDif} indicate that PR-derived HRTFs exhibit similar trends to those from high-resolution scans but with increased degradation. Both PR and 3D scan-based methods generally produce lower LSD values than a randomly selected non-individual HRTF across most of the frequency spectrum up to 12 kHz. However, significant discrepancies arise above 16 kHz, possibly due to errors in the BEM calculations as these appear for both reconstructions, with the PR one showing slightly larger deviations. As reported in \cite{hogg_hrtf_2024}, a randomly assigned non-individualised HRTF typically results in an average LSD of 5–10 dB across the spectrum.

Perceptual evaluation of the synthesised HRTFs is conducted through both localisation tasks and Spatial Release from Masking (SRM) experiments \cite{daugintis_effects_2024, gonzalez-toledo_spatial_2024}. The first assesses the ability of participants to accurately identify the direction of virtual sound sources presented at various azimuth and elevation angles. SRM experiments evaluate the impact of different HRTF sets on speech intelligibility in complex auditory scenes, asking participants to recognise target speech signals in the presence of spatially distributed maskers.

\subsection{Graph Neural Network}

The neural network developed in this study is based on a Graph Neural Network (GNN) architecture, where 3D meshes are represented as graphs in which vertices and edges encode geometry and connectivity data. Liu et al. introduced a novel framework for data-driven coarse-to-fine geometry modelling \cite{liu_neural_2020}, taking a coarse triangle mesh as input and recursively subdivides it to a finer geometry. This framework is adapted to our dataset to enhance the resolution of photogrammetry-reconstructed meshes.

A key challenge lies in the lack of a bijective map between the low- and high-resolution meshes, as they originate from different acquisition techniques. Unlike \cite{liu_neural_2020}, where a direct mapping exists, our approach leverages the method proposed by Schmidt et al. to compute continuous and bijective maps   (surface homeomorphisms) between genus-0 triangle meshes \cite{schmidt_surface_2023}. This requires adapting the dataset to meet the method’s constraints.

Once the bijective maps computed, the dataset of paired low- and high-resolution meshes trains a Graph Neural Network (GNN) to upscale low-resolution inputs to high-resolution outputs, optimised using a specific Hausdorff Distance-based loss function \cite{liu_neural_2020}. The model’s performance is validated by generating high-resolution meshes from unseen photogrammetry reconstructed meshes and evaluated geometrically with ground truth 3D scan meshes using the Hausdorff Distance.

\section{Expected Results and Challenges}

In photogrammetry reconstruction, the quality of the reconstructed mesh depends not only on the quality of the data but also on the algorithm and software used. Employing photogrammetry data taken at a fixed distance from the subject and limited on the horizontal plane, results in the reconstructed meshes to lack ear features and contain an incorrectly reconstructed overhead region. 

The numerical evaluation of HRTFs synthesised from raw PR meshes is expected to exhibit similar trends to those computed from high-resolution meshes, although with greater degradation due to geometric inaccuracies. Despite these limitations, perceptual evaluation through localisation tests is expected to demonstrate improved spatial accuracy compared to the KEMAR HRTF and approach the performance of acoustically measured HRTFs. 

On the model side, the GNN is expected to learn how to modify a mesh to upscale its resolution and improve ear morphology. However, a potential challenge lies in the risk of the GNN generalising ear morphology, which could lead to inaccurate fitting of personal anatomy. Future work is essential to better understand the perceptual relevance of various components of the pinna, as it remains unclear which parts of the ear are most crucial for accurate sound localisation and HRTF synthesis. By identifying and prioritising these key anatomical features, the GNN’s performance could be optimised for personal HRTF synthesis. The intended outcomes for the HRTFs synthesised from GNN-refined meshes are to achieve equivalent numerical and perceptual performance to those computed from high-resolution meshes. The data will be presented at the conference.

\section{Acknowledgments}
Horizon-MSCA-2022-DN-01: CherISH is a European Doctorate Network project funded by the European Union’s Horizon 2020 framework program for research and innovation under the Marie Sklodowska-Curie Grant Agreement No: 101120054.

% For bibtex users:
\bibliography{references}

% For non bibtex users:
%\begin{thebibliography}{citations}
%\bibitem{Author:00}
%E.~Author.
%\newblock The title of the conference paper.
%\newblock In {\em Proc.\ of the European Society on Vibration
%  }, pages 000--111, Chania, Greece, 2018.
%
%\bibitem{Someone:10}
%A.~Someone, B.~Someone, and C.~Someone.
%\newblock The title of the journal paper.
%\newblock {\em Acta Acust united Ac}, A(B):111--222, 2010.
%
%\bibitem{Someone:04}
%X.~Someone and Y.~Someone.
%\newblock {\em The Title of the Book}.
%\newblock S. Hirzel, Stuttgart, Germany, 2012.
%
%\end{thebibliography}

\end{document}